\documentclass{moriond}

\newcommand{\Photo}{} 
\usepackage{physics}
\usepackage{mathrsfs} 
\usepackage{bbold}

\newcommand{\cmb}{\mathscr{C}_2}

\begin{document}
\vspace*{4cm}
\title{Testing Bell inequalities and entanglement with di-boson final states}

\author{Luca Marzola}

\address{Laboratory of High Energy and Computational Physics, National Institute of Chemical Physics and Biophysics, R\"avala pst. 10, 10143 Tallinn, Estonia.}

\maketitle\abstracts{
We explore the phenomenology of quantum entanglement at collider experiments by computing the polarization density matrix of processes yielding two massive gauge bosons. After reviewing the formalism, we detail observables suitable to test the presence of entanglement and quantum correlations in the di-boson system. The implied violation of Bell inequalities can be observed with future data at the LHC in the decays of the Higgs boson to $Z$ boson pairs.}

\section{Quantum tomography}

Quantum tomography is a technique that aims to provide full knowledge of a quantum state by reconstructing the corresponding density matrix through suitable measurements. All quantities of interest can then be straightforwardly computed by applying the machinery of the density matrix formalism, including also observables that are not usually investigated in collider experiments. An important example is provided by the entanglement content of the quantum state under examination. Any interaction between two quantum systems, in fact, is bound to yield quantum correlations that entangle these systems, thereby bearing effects measurable with suitable quantum observables. Quantum correlations themselves can be subject to experimental investigations, aiming in this case to establish the nature of the theory underlying the processes we observe. Concerning this, John Bell showed in 1964~\cite{Bell1964} that the presence of quantum correlations makes it possible to distinguish between quantum mechanics and alternative classical theories proposing a local and deterministic description of Nature. The proof is encapsulated in an inequality that the correlations computed within local deterministic theories must respect. By evaluating the same correlations according to the rules of quantum mechanics, we find instead that the inequality can be violated. 

In the following\footnote{Drawing from Refs.~\cite{Fabbrichesi:2023cev,Fabbrichesi:2023jep} prepared in collaboration with M. Fabbrichesi, R. Floreanini and E. Gabrielli.} we discuss some of the possibilities offered by quantum tomography at collider experiments focusing mostly on $ZZ$ states. Given that $W$ and $Z$ bosons have three polarization modes each, we can regard these objects as quantum three-level systems -- qutrits, in short -- and investigate correlations in the polarization space spanned by two qutrits originated in an interaction.

\section{Qutrits} 
The transition amplitude leading to the production of a massive gauge boson with polarization $\lambda\in\{+1, 0, -1\}$ and momentum $p$ can be written as
\begin{equation}
    \mathcal{M}(\lambda, p) = \mathcal{A}_\mu \varepsilon^{\mu*}_\lambda(p)
    \label{eq:amp} 
\end{equation}
and the state $\ket{V^\mu}$ of the boson $V$ is consequently determined as
\begin{equation}
    \ket{V^\nu} = \sum_\lambda \mathcal{M}(\lambda) \varepsilon^{\nu}_\lambda.
    \label{eq:state}
\end{equation}
To construct the corresponding (covariant) density matrix we proceed as usual, obtaining
\begin{equation}
    \rho^{\mu\nu} = - \frac{\dyad{V^\mu}{V^\nu}}{\braket{V^\mu}{V_\mu}} 
    \label{eq:rcov}
\end{equation}
after the normalization of the state vector and having inserted a factor of (-1) to account for the signature (1, -1, -1, -1) of the Minkowski metric $g_{\mu\nu}$. To obtain the polarization density matrix we then move to polarization space by use of the mapping $\mathcal{P}^{\mu\nu}_{\lambda\lambda'}(p)=\varepsilon^{\mu*}_\lambda(p) \, \varepsilon^{\nu}_{\lambda'}(p) \, $:
\begin{equation}
    \rho_{\lambda\lambda'}=\mathcal{P}^{\mu\nu}_{\lambda\lambda'}\, \rho_{\mu\nu}.
    \label{eq:rpol}
\end{equation}
It follows from the orthonormality relation $g_{\mu\nu}\,\varepsilon^\mu_\lambda(p) \,\varepsilon^\nu_{\lambda'}(p)=-\delta_{\lambda\lambda'}$ and Eqs.~\eqref{eq:amp}-\eqref{eq:rpol} that
\begin{equation}
    \rho_{\lambda\lambda'}=\frac{\mathcal{M}(\lambda)\mathcal{M}^\dagger(\lambda')}{\sum_{\lambda''} \mathcal{M}^\dagger(\lambda'')\mathcal{M}(\lambda'')}
    =\frac{\mathcal{A}_\mu \mathcal{A}^\dagger_\nu \mathcal{P}^{\mu\nu}_{\lambda\lambda'}}{\abs{\bar{\mathcal{M}}}^2}.
    \label{eq:rpol2}
\end{equation}  
In order to compute the polarization density matrix from the amplitude of the underlying process we then need an expression for the mapping $\mathcal{P}$. To this end, consider the explicit form of the wave vector of our massive gauge boson
\begin{equation}
    \varepsilon^{\mu}_\lambda (p)=-\frac{1}{\sqrt{2}}|\lambda|\left(\lambda \, n_1^{\mu}+i \, n_2^{\mu}\right)
+\Big(1-|\lambda| \Big)n_3^{\mu} ,
\label{eps}
\end{equation}
where the four-vectors $n_i=n_i(p)$, $i\in\{1,2,3\}$, are obtained by boosting the linear polarization vectors defined in the frame where the boson is at rest -- their spatial components form a right-handed system -- to a frame where it has momentum $p$. With the above expression we find~\cite{Kim1980,Choi1989}
\begin{equation}
\mathcal{P}^{\mu\nu}_{\lambda \lambda '}(p) =
\frac{1}{3}\left(-g^{\mu\nu}+\frac{p^{\mu}p^{\nu}}{m_V^2}\right)
\delta_{\lambda\lambda '}-\frac{i}{2m_V}
\epsilon^{\mu\nu\alpha\beta}p_{\alpha} n_{i\,\beta} \left(S_i\right)_{\lambda\lambda '}-\frac{1}{2}n_i^{\mu}n_j^{\nu} \left(S_{ij}\right)_{\lambda\lambda '},
\label{proj}
\end{equation}
where $m_V$ is the invariant mass of the vector boson $V$, $\epsilon^{\mu\nu\alpha\beta}$ the permutation symbol ($\epsilon^{0123}=1$) and $S_i$, $i\in\{1,2,3\}$, are the $SU(2)$ generators in the spin-1 representation --  the eigenvectors of $S_3$, corresponding to the eigenvalues $\lambda\in\{+1, 0, -1\}$, define the helicity basis. The spin matrix combinations appearing in the last term are given by
\begin{equation}
S_{ij}= S_iS_j+S_jS_i-\frac{4}{3} \mathbb{1}\, \delta_{ij},
\label{Sij}
\end{equation}
with $i,j\in\{1,2,3\}$ and $\mathbb{1}$ being the $3\times 3$ unit matrix.

Eqs.~\eqref{eq:rpol2} and \eqref{proj} make it possible to compute the polarization density matrix for an ensemble of $V$ bosons produced in repeated reactions described by the amplitude $\mathcal{M}$. The formalism can be straightforwardly extended to processes yielding a bipartite qutrit state formed by two massive gauge bosons, $V_1$ and $V_2$. In this case we have
\begin{equation}
    \rho=\frac{{\mathcal A}_{\mu\nu}{\mathcal A}^{\dagger}_{\mu'\nu'}}{\abs{\bar{\mathcal{M}}}^2}
    \left[ \mathcal{P}^{\mu\mu'}(k_1)\otimes\mathcal{P}^{\nu\nu'}(k_2) \right],
    \label{eq:rhoVV}
\end{equation}
where $k_1$ and $k_2$ denote the momenta of the vector bosons in a given frame (the center of mass frame in the following) and the tensor product, $\otimes$, operates in polarization space producing a $9\times9$ polarization density matrix.  

From the experimental side, information about the polarization of decaying $W$ and $Z$ bosons is carried by the directions of the emitted charged leptons. The polarization density matrix of the system of interest can then be reconstructed from collider data by analyzing the angular distribution of the decay products of the massive gauge bosons, as illustrated in Refs.~\cite{Ashby-Pickering:2022umy,Fabbrichesi:2023cev} for a bipartite qutrit state. Alternatively, the density matrix can be obtained from measurements of the helicity amplitudes of the underlying production process, as shown for instance in Ref.~\cite{Fabbrichesi:2023jep} for a simple case in which the bipartite qutrit state is pure ($\Tr(\rho^2)=1$).   

\section{Quantum observables} 
Before proceeding with the analysis, we detail the observables suitable to explore the entanglement content of a system and to assess the possible violation of Bell inequalities. 

\subsection{Entanglement}
The problem of quantifying the entanglement content of a quantum state is highly non-trivial~\cite{horodecki2009quantum}. In the simple case provided by bipartite pure states, the entropy of entanglement 
\begin{equation}
    \mathscr{E}[\rho] = - \Tr[\rho_A \log \rho_A] =  - \Tr[\rho_B \log \rho_B]\label{E}
\end{equation}
provides a genuine entanglement measure. The reduced density matrices $\rho_A=\Tr_B(\rho)$ and $\rho_B=\Tr_A(\rho)$ are obtained through the partial trace of the bipartite system density matrix $\rho$ over the degrees of freedom of the component subsystem $B$ and $A$, respectively. The entropy of entanglement then amounts to the von Neumann entropy of either subsystem and satisfies $0\leq  \mathscr{E}[\rho]\leq \ln 3$ for a two-qutrit state. Vanishing values signal a separable state: a state with no quantum correlations entwining its subsystems and, therefore, \textit{not} entangled. The maximum entropy value is achieved instead by states that are maximally entangled.

For bipartite systems which are not pure, or for mixed states, we can only rely on entanglement witnesses meant to provide conditions sufficient to establish the presence of entanglement. To see an example, consider a bipartite pure state $\ket{\Psi}$ with components $A$ and $B$. Its concurrence is defined as~\cite{rungta2001universal}
\begin{equation}
    {\cal C}[\ket{\Psi}]=\sqrt{2\left( 1-{\rm Tr}\big[(\rho_r)^2\big]\right)}, \quad r\in\{A,B\},
    \label{C_psi}
\end{equation}
and vanishes if and only if the state is separable. For a mixed state described on a basis of pure states $\{\ket{\Psi_i}\}$ by a density matrix $\rho=\sum_i p_i\, \dyad{\Psi_i}{\Psi_i}$, with $p_i>0$ and $\sum_i p_i=1$,  the definition is generalized in its convex roof extension
\begin{equation}
{\cal C}[\rho]=\underset{\{\ket{\Psi}\}}{\rm inf} \sum_i p_i\, {\cal C}[\ket{\Psi}],
\label{C_rho}
\end{equation}
where the infimum is obtained by considering all the possible decompositions of $\rho$ into different sets of pure states. As the concurrence vanishes also for mixed states that are separable~\cite{mintert2004concurrence}, the quantity can serve as an entanglement detector. Unfortunately, the optimization problem posed by Eq.~\eqref{C_rho} makes the computation of the concurrence so demanding that decent analytical solutions are known only for the case of mixed bipartite two-level systems. Lower bounds on ${\cal C}$ are more manageable, in particular~\cite{mintert2007observable} $\cmb[\rho] \leq \left({\cal C}[\rho]\right)^{2}$ can be computed for a two-qutrit system as
\begin{equation}
    \cmb [\rho] = 2 \,\text{max}\, \Big( 0,\, \Tr[\rho^{2}] - \Tr[(\rho_A)^2],\, \Tr[\rho^{2}] - \Tr[(\rho_B)^2]  \Big),
\label{C_bound}
\end{equation} 
where again $\rho_A$ and $\rho_B$ are the reduced density matrices of the qutrits. A non-vanishing value of $\cmb$ implies a non-vanishing concurrence, thus witnessing the presence of entanglement in the system.

\subsection{Bell inequalities}

To discriminate between quantum mechanics and alternative local deterministic theories, we use an instance of the Bell inequality optimized for qutrits: the CGLMP inequality~\cite{collins2002bell,kaszlikowski2002clauser}. The test defines the observable
\begin{align}
    {\cal I}_{3}& = P(A_{1}=B_{1}) +  P(B_{1}=A_{2}+1) +  P(A_{2}=B_{2}) +  P(B_{2}=A_{1}) \nonumber \\
&  -P(A_{1}=B_{1}-1) - P(A_{1}=B_{2}) -P(A_{2}=B_{2}-1) -P(B_{2}=A_{1}-1) 
\label{CGLMP}
\end{align}
in terms of the probabilities $P(A_i - B_j +k)$ that the outcomes $(0,1$ or $2)$ of the operators $\hat A_{1,2}$ and $\hat B_{1,2}$, acting on the homonymous qutrits, differ by $k$ modulo 3. For classical local and deterministic theories it holds $ {\cal I}_{3}\leq 2$. The bound can be instead violated if the above probabilities are obtained through the rules of quantum mechanics  
\begin{equation}
    {\cal I}_{3}  = \Tr[\rho \mathcal{B}],
    \label{eq:I3}
\end{equation}
where the Bell operator $\mathcal{B}$ depends on the choice of operators used in the test. To maximize the effect we use~\cite{acin2002quantum}
\begin{equation}
    {\cal B} = \left( \begin{smallmatrix} 
        0 & 0 & 0 & 0 & 0 & 0 & 0 & 0 & 0  \\
        0 & 0 & 0 & -\frac{2}{\sqrt{3}} & 0 & 0& 0 & 0 & 0  \\
        0 & 0 &0 & 0 & -\frac{2}{\sqrt{3}} & 0 &2 & 0 & 0  \\
        0 &  -\frac{2}{\sqrt{3}} & 0 & 0 & 0 & 0 & 0 &0 & 0  \\
        0& 0 & -\frac{2}{\sqrt{3}} & 0 & 0 & 0 & -\frac{2}{\sqrt{3}} & 0 &0  \\
        0 & 0 & 0 & 0 & 0 & 0 & 0 &  -\frac{2}{\sqrt{3}} & 0  \\
        0 & 0 & 2 & 0 & -\frac{2}{\sqrt{3}} & 0 &0& 0 & 0  \\
        0 & 0 & 0 & 0 & 0 &  -\frac{2}{\sqrt{3}} & 0 & 0 & 0  \\
        0 & 0 & 0 & 0 & 0& 0 & 0 & 0 & 0  \\
      \end{smallmatrix} \right)
      \label{eq:Bo}
\end{equation}
and further optimize the observable by performing local unitary transformations ${\cal B} \to (U\otimes V)^{\dag} \cdot {\cal B}\cdot (U\otimes V)$ independently for each of the considered kinematic configurations.

\section{Qutrits at colliders}
We begin by investigating entanglement and the prospects for detecting Bell inequalities violation at colliders in the simple case given by di-boson production in Higgs boson decays. 

\subsection{A first example: $H\to ZZ^*$}
Consider a pair of $Z$ bosons emitted in the resonant scattering $pp\to H\to ZZ^*$, where a star denotes an off-shell particle. We treat the latter as an on-shell state with a mass reduced by a factor $f<1$: $M_{Z}^*= f M_Z$. The polarization density matrix obtained for this process with Eq.~\eqref{eq:rhoVV} is that of a pure state: a coherent superposition of the $m=0$ components belonging to the $J=0,1$ and $2$ multiplets formed by the polarizations of the two $Z$ bosons. We can then write  
\begin{equation}
    \rho_{ZZ^*} = \dyad{\Psi_{ZZ^*}}{\Psi_{ZZ^*}},
    \qquad
    \ket{\Psi_{ZZ^*}} = \frac{1}{\sqrt{2 +  \kappa^2}} \left[  \ket{+1}\otimes\ket{-1} -  \kappa \ket{0}\otimes\ket{0} + \ket{-1}\otimes\ket{+1} \right] \label{pure}   
\end{equation}
where $\kappa = 1+ [m_H^2 - (1+f)^2 M_Z^2]/(2 f M^2_Z)$ and $m_H$ is the Higgs boson mass.
\begin{figure}[h!]
\begin{center}
    \includegraphics[width=0.45\linewidth]{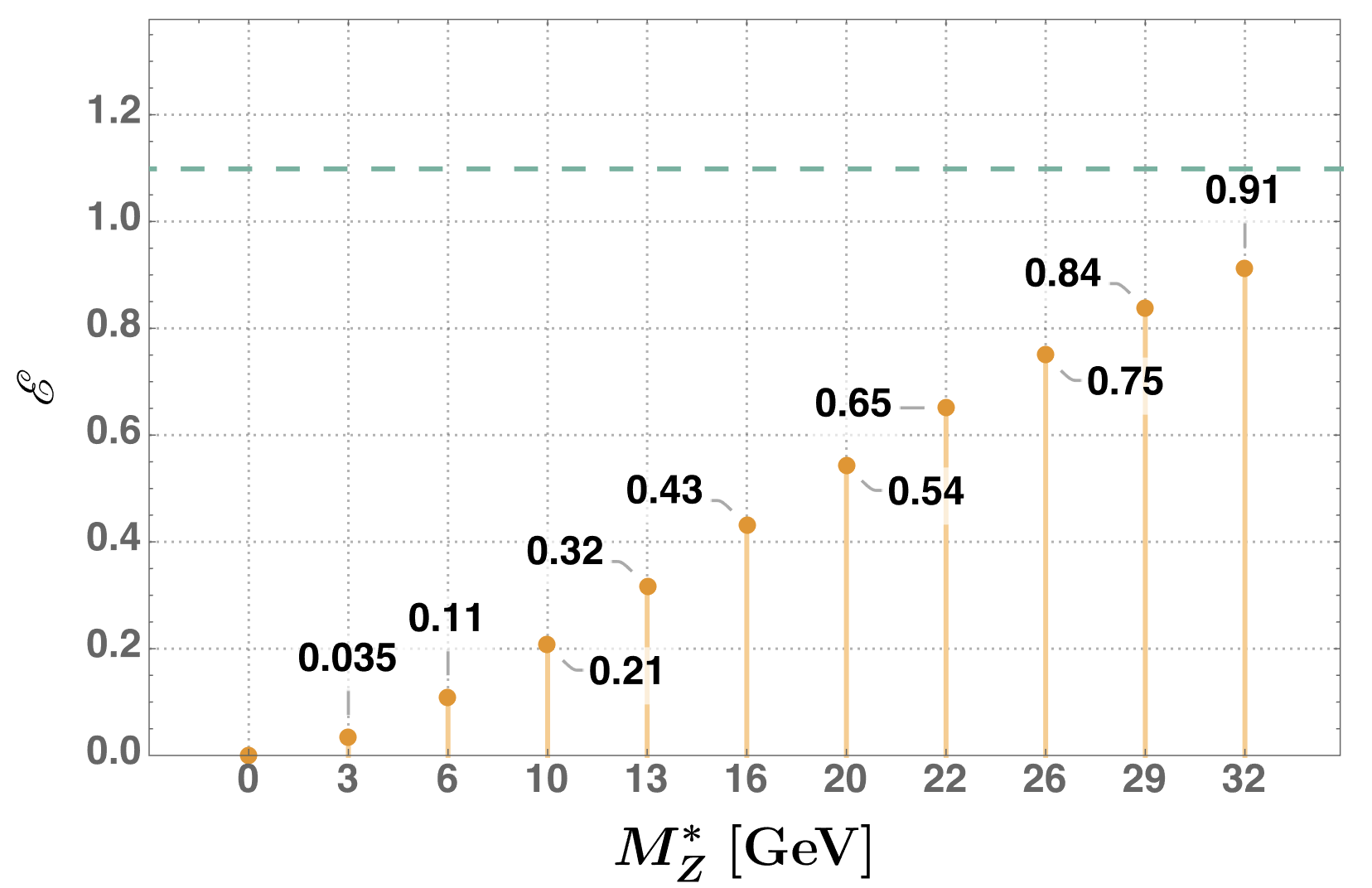}
    \includegraphics[width=0.437\linewidth]{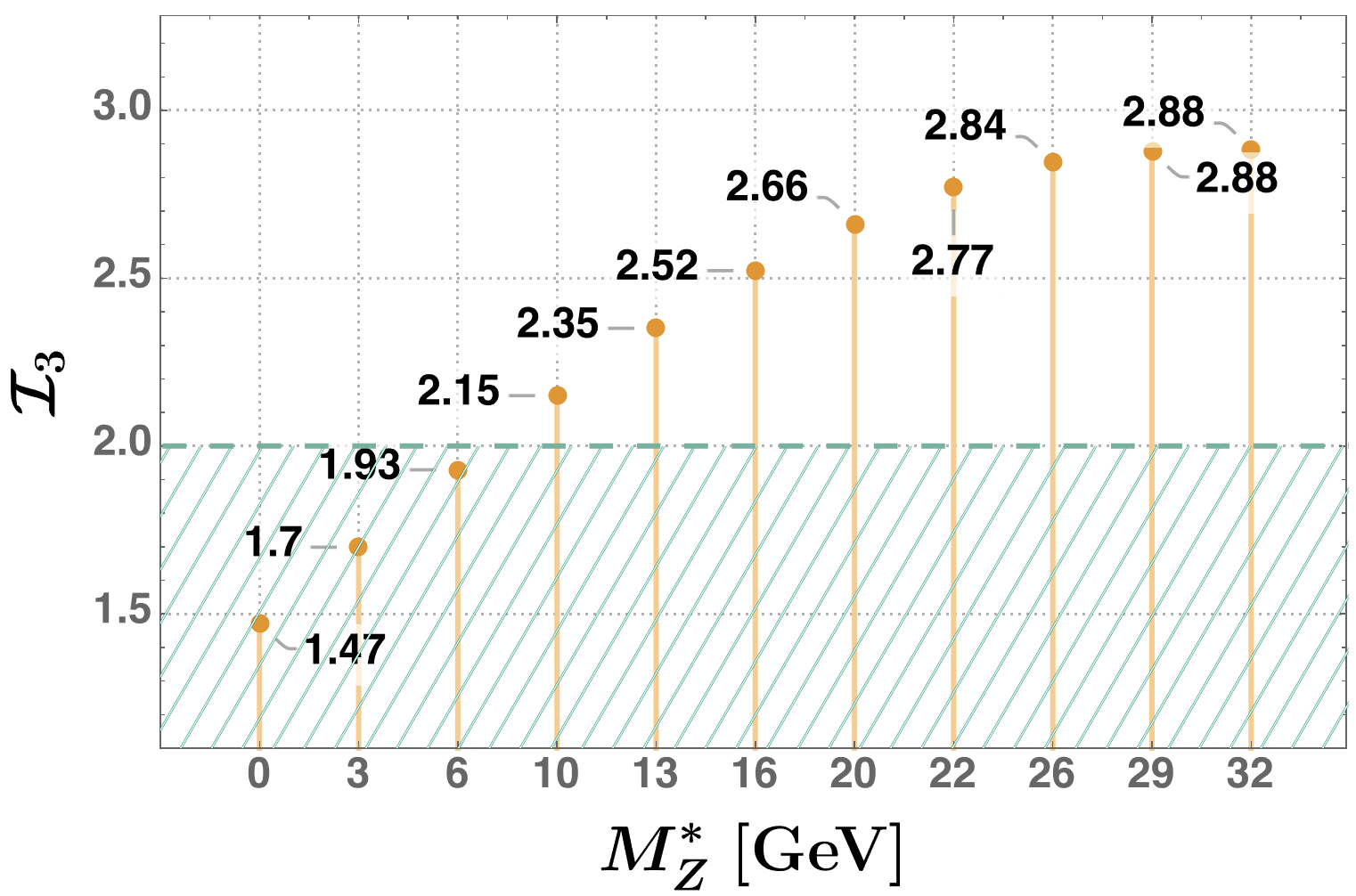}
    \caption{\label{fig:HZZ} Entanglement content of the $\ket{\Psi_{ZZ^*}}$ state produced in $H\to ZZ^*$ decays (left panel) and expectation value of the Bell operator~\eqref{eq:I3} (right panel) as functions of the virtual $Z$ boson mass. }
\end{center}
\end{figure}
Fig.~\eqref{fig:HZZ} shows the results obtained for the entropy of entanglement~\eqref{E} and Bell inequality violation~\eqref{eq:I3} as functions of the virtual $Z$ boson mass $M^*_Z$. As we can see in the first panel, the entropy grows with the parameter and it can be shown to reach the maximal allowed value (indicated by the dashed line) at the kinematic threshold of the decay process, where the qutrits combine into a singlet state ($\kappa=1$). The falling behavior is due to the circular helicity components in~\eqref{pure} being progressively suppressed as we let $f\to 0$ ($\kappa\gg1$). The second panel of Fig.~\eqref{fig:HZZ} shows instead that Bell inequalities are generally violated in the analyzed process (${\cal I}_3 >2$), confirming the results of numerical simulations~\cite{Aguilar-Saavedra:2022wam,BARR2022136866}. To quantify the power of the test, we run 1000 pseudo-experiments changing the virtual boson mass and considering statistical errors driven by the number of suitable events available at the LHC run2 and Hi-Lumi phase. The significances thus obtained are 1.3 and 5.6, respectively, assuming a 70\% efficiency in the reconstruction of final state leptons. 

Similar results would hold for the $WW$ channel if systematic uncertainties pertaining to the reconstruction of neutrino momenta were not to dominate the error balance.

\subsection{A second example: $pp\to ZZ$}
\begin{figure}[h!]
    \begin{center}
        \includegraphics[width=0.42\linewidth]{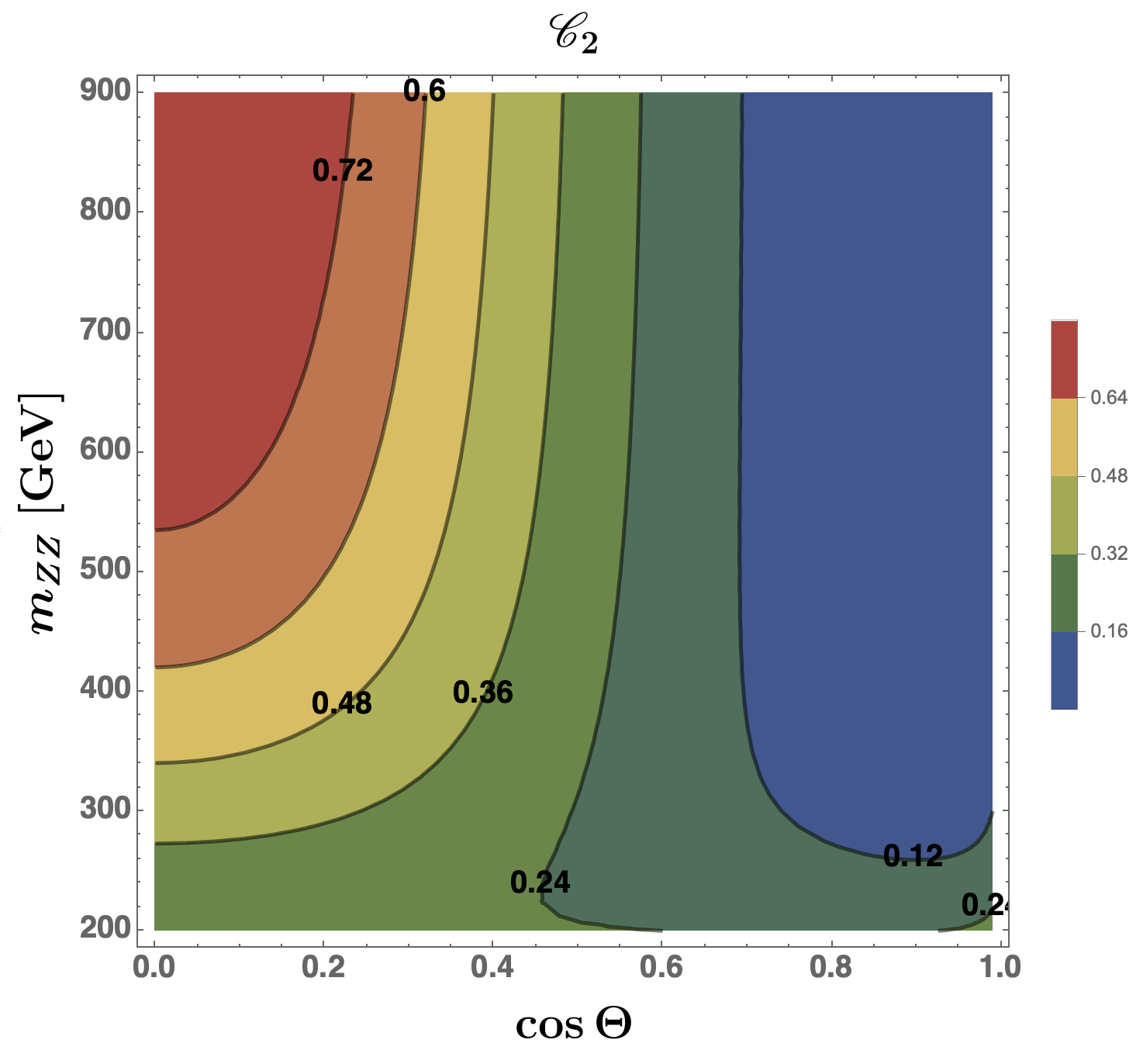}
        \hspace{0.2cm}
        \includegraphics[width=0.42\linewidth]{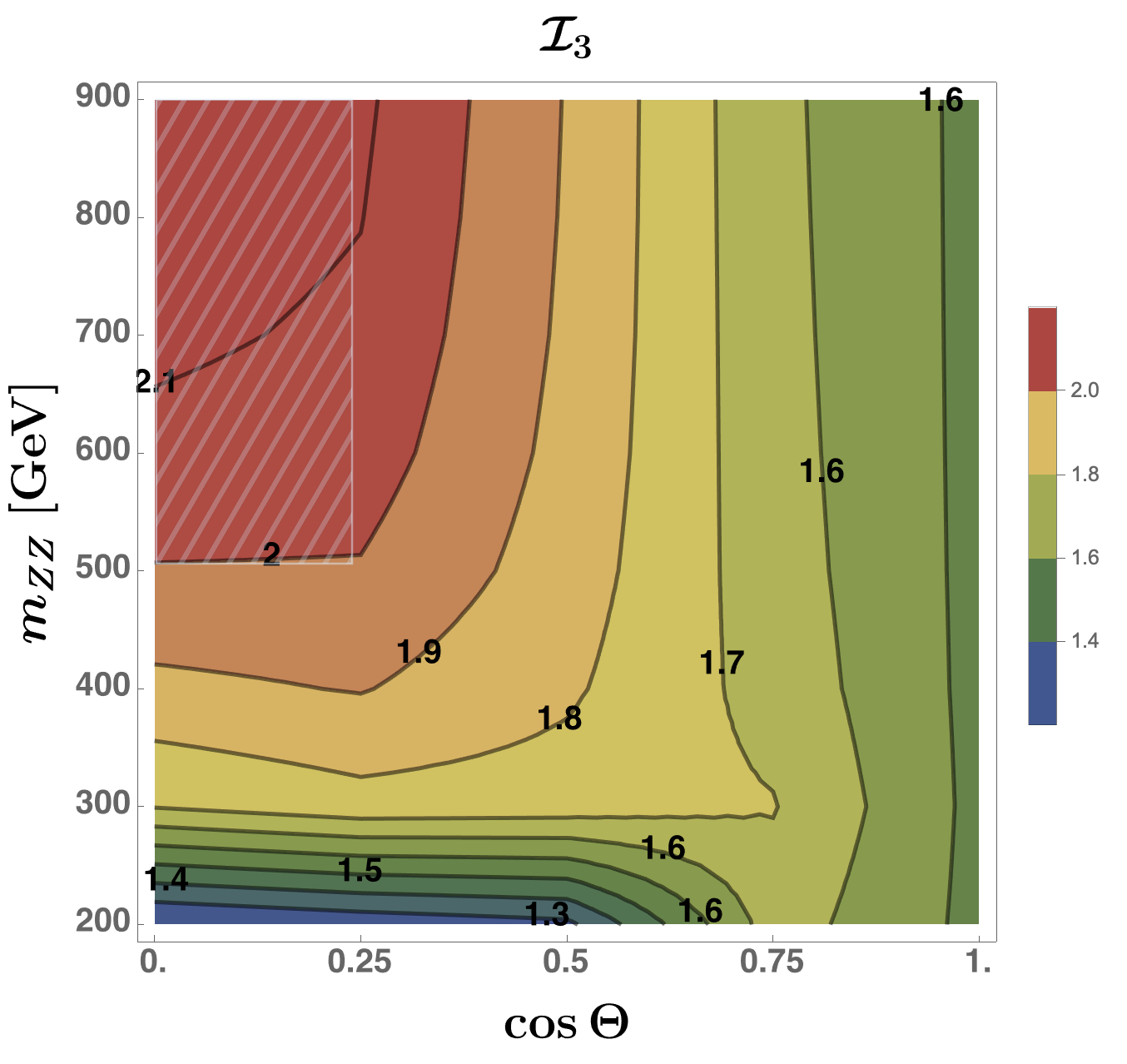}
        \caption{\label{fig:ppZZ} Minimum entanglement of a $Z$ boson pair produced via quark fusion in $pp$ collisions (left panel) and expectation value of the Bell operator~\eqref{eq:I3} (right panel) as functions of the $ZZ$ invariant mass and center of mass scattering angle.}
    \end{center}
\end{figure}

As a second example we focus on the production of two real $Z$ bosons via quark fusion in $pp$ collisions. The polarization density matrix $\rho_{ZZ}$ is given in this case by the convex combination of the density matrices $\rho^{q_1 \bar q_2}_{ZZ}$ for the parton level production modes
\begin{equation}
  \rho_{ZZ} = \sum_{\{q_1 \bar q_2\}} w^{q_1\bar q_2} \, \rho^{q_1 \bar q_2}_{ZZ}, 
  \qquad
  w^{q_1 \bar q_2} 
  = 
  \frac{L^{q_1\bar q_1} \, \abs{\bar{\mathcal{M}}^{q_1 \bar q_2}_{ZZ}}^2 }
  {\sum_{\{q_1 \bar q_2\}} L^{q_1\bar q_1} \, \abs{\bar{\mathcal{M}}^{q_1 \bar q_2}_{ZZ}}^2}.
  \label{eq:rhopzz}
\end{equation}
The sums run here over all the possible initial quark configurations and $\abs{\bar{\mathcal{M}}^{q_1 \bar q_2}_{ZZ}}^2$ is the squared amplitude of the parton process summed over the degrees of freedom of the initial state. The weights of the convex combination depend also on the parton luminosities $L^{q_1\bar q_1} $ specific to the initial states $q_1\bar{q_2}$. The left panel of Fig.~\ref{fig:ppZZ} shows our results for the presence of entanglement in the polarizations of the two $Z$ bosons, obtained with the quantum witness $\cmb[\rho]$ as the density matrix in Eq.~\eqref{eq:rhopzz} describes a mixture of $ZZ$ states. The entanglement, revealed by $\cmb[\rho]>0$, is stronger in the kinematical region characterized by low values of the center of mass scattering angle and large invariant di-boson mass $m_{ZZ}$. 
Our results concerning Bell inequalities follow roughly the same pattern, leading to an observable violation in correspondence of the kinematical configurations under the hatched area in the right panel of Fig.~\ref{fig:ppZZ}. To assess the power of the test we compute the relevant cross-section corrected by NNLO $k$-factors and impose kinematic cuts to retain only events in the selected signal area. The statistical error is modelled by using the number of suitable events available at the LHC, after considering a 70\% efficiency in the identification of each final state lepton. Performing $10^4$ pseudo-experiments, obtained by varying the scattering angle and the di-boson invariant mass in the ranges of interest, we find that run2 data give an average value of the observable below the required threshold: ${\cal I}_{3}\leq 2$. Differently, the significance expected with Hi-Lumi data is more than 2. 

A similar analysis for the competing $WW$ channel achieves a significance of $0.8$ with run2 data, but the result does not improve with the Hi-Lumi projections because of the large systematic error due to final state neutrinos.   

\vspace{0.25cm}
It is our hope that the presented examples could encourage the LHC  collaborations to perform simulations dedicated to the assessment of the discussed quantum observables. Eventually, these could be included as additional observables in the routine data analyses performed.

\section*{Acknowledgments} 

This work was supported by the Estonian Research Council grant PRG356.

\section*{References}
 
\bibliography{lucamarzola.bib}

\begin{thebibliography}{10}

\bibitem{Bell1964}
J.~S. Bell.
\newblock {On the Einstein Podolsky Rosen paradox}.
\newblock {\em Physics}, 1:195--200, 1964.

\bibitem{Fabbrichesi:2023cev}
M.~Fabbrichesi, R.~Floreanini, E.~Gabrielli, and L.~Marzola.
\newblock {Bell inequalities and quantum entanglement in weak gauge bosons
  production at the LHC and future colliders}.
\newblock 2 2023.

\bibitem{Fabbrichesi:2023jep}
M.~Fabbrichesi, R.~Floreanini, E.~Gabrielli, and L.~Marzola.
\newblock {Stringent bounds on $HWW$ and $HZZ$ anomalous couplings with quantum
  tomography at the LHC}.
\newblock 4 2023.

\bibitem{Kim1980}
J.~Kim, J.~E. Kim, and H.~S. Song.
\newblock A consistent way of treating arbitrary spin polarization.
\newblock 1980.
\newblock PRINT-80-0470 (SEOUL-NATIONAL).

\bibitem{Choi1989}
S.~Y. Choi, Taeyeon Lee, and H.~S. Song.
\newblock Density matrix for polarization of high-spin particles.
\newblock {\em Phys. Rev. D}, 40:2477--2480, Oct 1989.

\bibitem{Ashby-Pickering:2022umy}
Rachel Ashby-Pickering, Alan~J. Barr, and Agnieszka Wierzchucka.
\newblock {Quantum state tomography, entanglement detection and Bell violation
  prospects in weak decays of massive particles}.
\newblock {\em JHEP}, 05:020, 2023.

\bibitem{horodecki2009quantum}
Ryszard Horodecki, Pawe{\l} Horodecki, Micha{\l} Horodecki, and Karol
  Horodecki.
\newblock Quantum entanglement.
\newblock {\em Reviews of Modern Physics}, 81(2):865--942, 2009.

\bibitem{rungta2001universal}
Pranaw Rungta, V~Bu{\v{z}}ek, Carlton~M Caves, M~Hillery, and GJ~Milburn.
\newblock Universal state inversion and concurrence in arbitrary dimensions.
\newblock {\em Physical Review A}, 64(4):042315, 2001.

\bibitem{mintert2004concurrence}
Florian Mintert, Marek Ku{\'s}, and Andreas Buchleitner.
\newblock Concurrence of mixed bipartite quantum states in arbitrary
  dimensions.
\newblock {\em Physical Review Letters}, 92(16):167902, 2004.

\bibitem{mintert2007observable}
Florian Mintert and Andreas Buchleitner.
\newblock Observable entanglement measure for mixed quantum states.
\newblock {\em Physical Review Letters}, 98(14):140505, 2007.

\bibitem{collins2002bell}
Daniel Collins, Nicolas Gisin, Noah Linden, Serge Massar, and Sandu Popescu.
\newblock Bell inequalities for arbitrarily high-dimensional systems.
\newblock {\em Physical Review Letters}, 88(4):040404, 2002.

\bibitem{kaszlikowski2002clauser}
Dagomir Kaszlikowski, L~C Kwek, Jing-Ling Chen, Marek Zukowski, and C~H Oh.
\newblock Clauser-horne inequality for three-state systems.
\newblock {\em Physical Review A}, 65(3):032118, 2002.

\bibitem{acin2002quantum}
A~Acin, T~Durt, N~Gisin, and J~I Latorre.
\newblock Quantum nonlocality in two three-level systems.
\newblock {\em Physical Review A}, 65(5):052325, 2002.

\bibitem{Aguilar-Saavedra:2022wam}
J.~A. Aguilar-Saavedra, A.~Bernal, J.~A. Casas, and J.~M. Moreno.
\newblock {Testing entanglement and Bell inequalities in $H\to ZZ$}.
\newblock {\em Phys. Rev. D}, 107(1):016012, 2023.

\bibitem{BARR2022136866}
Alan~J. Barr.
\newblock Testing bell inequalities in higgs boson decays.
\newblock {\em Physics Letters B}, 825:136866, 2022.

\end{thebibliography}
\end{document}